# "Spectral implementation" for creating a labeled pseudo-pure state and the Bernstein–Vazirani algorithm in a four-qubit nuclear magnetic resonance quantum processor

Xinhua Peng,[a)] Xiwen Zhu, Ximing Fang,[b)] Mang Feng, Maili Liu, and Kelin Gao
*Laboratory of Magnetic Resonance and Molecular Physics, Wuhan Institute of Physics and Mathematics, the Chinese Academy of Sciences, Wuhan 430071, People's Republic of China*

A quantum circuit is introduced to describe the preparation of a labeled pseudo-pure state by multiplet-component excitation scheme which has been experimentally implemented on a 4-qubit nuclear magnetic resonance quantum processor. Meanwhile, we theoretically analyze and numerically investigate the low-power selective single-pulse implementation of a controlled-rotation gate, which manifests its validity in our experiment. Based on the labeled pseudo-pure state prepared, a 3-qubit Bernstein–Vazirani algorithm has been experimentally demonstrated by spectral implementation. The "answers" of the computations are identified from the split peak positions in the spectra of the observer spin, which are equivalent to projective measurements required by the algorithms. © *2004 American Institute of Physics.*
[DOI: 10.1063/1.1642579]

## I. INTRODUCTION

Quantum mechanics provides a novel way to process information that extends the extraordinary capabilities of a quantum processor beyond those available on a classical physical system.[1–3] Some problems are found that are able to be solved exponentially faster on a quantum computer than on a classical computer, e.g., factoring[4] and quantum simulation.[5] Other problems, such as those to analyze the contents of a "black box" *oracle* that performs *a priori* unknown unitary transformation, can be solved exponentially or polynomially faster relative to the oracle[6] if feeding quantum superposition to the box. These algorithms include Grover's algorithm,[7] Deutsch's algorithm,[8] Deutsch–Jozsa algorithm,[9] Simon's algorithm,[10] and the Bernstein–Vazirani algorithm,[11] etc. Hitherto, many possible approaches have been proposed for the physical implementation of quantum computers.[3] Of these methods, liquid-state nuclear magnetic resonance (NMR) is arguably the most successful one[12] and has established small (up to 7 quantum bits) but fully operational quantum processors.[13] These experiences of implementing liquid-state NMR quantum computation may be useful to build up a robust and practical quantum computer. The recent research showed that NMR techniques were successfully applied to ion trap quantum computing.[14]

Different from the usual scheme of energy-level implementation, Madi *et al.* proposed "spectral implementation of a quantum computer"[15] by employing an additional observer spin $I_0$ that is coupled to the spins carrying $n$ computational quantum bits (qubits) $I_1, I_2, \ldots, I_n$, which allows a mapping of the states of a quantum computer on a set of transitions between energy levels. Concretely, $2^n$ logic states of an $n$-qubit quantum computer are assigned to $2^n$ individual spectral resonance lines of spin $I_0$, each of which can be conveniently represented by the operators in spin Liouville space

$$I_{0x} I_1^{\alpha/\beta} I_2^{\alpha/\beta} \cdots I_n^{\alpha/\beta}, \tag{1}$$

where $I_i^\alpha = |0\rangle_{ii}\langle 0| = \frac{1}{2}(E_i + 2I_{iz})$, $I_i^\beta = |1\rangle_{ii}\langle 1| = \frac{1}{2}(E_i - 2I_{iz})$, $I_i^\alpha + I_i^\beta = E_i$ (the 2×2 unit matrix), $2I_{i\eta} = \sigma_{i\eta}$ ($\eta = x, y, z$) are Pauli operators and subscript $i$ denotes the $i$th spin. Each of qubits $I_1, I_2, \ldots, I_n$ is either in the state $|0\rangle$ or $|1\rangle$. Assuming that spin $I_0$ has resolved scalar $J$ couplings to all other $n$ qubits $I_1, I_2, \ldots, I_n$, all logic states of the $n$-qubit quantum computer are distinguishable due to the nondegenerate single-quantum transitions of spin $I_0$. Hence, instead of a standard pseudo-pure state (PPS) with deviation $I_1^\alpha I_2^\alpha \cdots I_n^\alpha$ on $n$ spin-1/2 nuclei, one can prepare a so-called "labeled" PPS (LPPS) with deviation $I_{0z} I_1^\alpha I_2^\alpha \cdots I_n^\alpha$ on $n+1$ spins as an initial state of computation by employing an additional observer spin $I_0$. After a single spin-selective pulse $(\pi/2)_y^0$, a LPPS is transferred to one of Eq. (1), which is easily recognizable in a NMR spectrum of spin $I_0$, i.e., only one of the peaks labeled by the logic state $|00\cdots 0\rangle$ is visible. An algorithmic benchmark for quantum information processing has been demonstrated in a liquid-state NMR system by creating such a LPPS.[16,17] Moreover, the LPPS can be used in error-correcting codes.[18] As the final answer of a computation is possibly achieved by only detecting the observer spin $I_0$, "spectral implementation" scheme of a quantum computer provides another way to read out the answers of certain computational problems, such as database search, and appears advantageous compared to a mapping on the energy levels themselves.[15]

---

[a)]Also at: Department of Physics, University of Dortmund, D-44221 Germany; Electronic mail: xhpeng@wipm.ac.cn and xinhua@e3.physik.uni-dortmund.de
[b)]Also at: Department of Physics, Hunan Normal University, Changsha 410081, People's Republic of China.

In this paper, we introduce a quantum circuit to describe Madi's multiplet-component excitation scheme of creating a LPPS[15] which has been experimentally implemented on a 4-spin sample. Further, by considering the exact effective Hamiltonian in a rotating frame for a low-power radio-frequency (RF) pulse on a single multiplet-component (transition-selective excitation, TSE) in certain week-coupled spin system, we theoretically analyze that a multi-qubit controlled-rotation gate can be approximately obtained up to conditional phases factors whenever the RF field power is low compared to the spin–spin coupling. The gate fidelities are numerically simulated for different rotating angles and RF field powers on the 3-spin and 4-spin systems, which theoretically verifies the feasibility of implementing a single-pulse quantum controlled-rotating gate by a suitable choice of the RF field power on specific initial states. Besides, a 3-qubit quantum algorithm to solve the Bernstein–Vazirani parity problem has been experimentally investigated on such a NMR quantum processor. The reading-out step consists of a single spin-selective pulse followed by acquisition of the signal. The answer of the computation is encoded in the amplitudes and signs of various multiplet components in the spectrum of the observer spin.

## II. CREATION OF A LABELED PSEUDO-PURE STATE

The preparation of a proper initial state is one of the most important requirements for a useful computation. Starting from thermal equilibrium, many methods have been used to prepare a standard PPS in NMR, including spatial averaging,[19–21] temporal averaging[22,23] and logical labeling.[24–26] For a LPPS such as $I_{0z}I_1^\alpha I_2^\alpha \cdots I_n^\alpha$, several methods have also been proposed such as Madi's multiplet-component excitation scheme,[15] the cat-state benchmark,[16] spatially encoding[27] and generally spatial averaging.[28] Here, we first introduce a quantum circuit to describe Madi's multiplet-component excitation scheme.

### A. Quantum circuit of creating a LPPS

Consider a system with $n+1$ spins $I=1/2$ (for both a homonuclear and heteronuclear system), whose deviation density matrix of thermal equilibrium under the high temperature approximation can be written as

$$\rho_{eq} = \sum_{i=0}^{n} \omega_i I_{iz} \qquad (2)$$

with $\omega_i$ being the Larmor frequency of spin $I_i$. The state $I_{0z}I_1^\alpha I_2^\alpha \cdots I_n^\alpha$ can be achieved from thermal equilibrium $\rho_{eq}$ by applying a nonselective $(\pi/2)_y$ pulse on $n+1$ spins and a transition-selective $(\pi/2)_{-y}^k$ pulse on a single multiplet line of spin $I_0$, for example, on the transition where all spins $I_1, I_2, \ldots, I_n$ are in their $|0\rangle$ states, followed by a pulsed-field gradient (PFG).[15] Any of the other states $I_{0z}I_1^{\alpha/\beta}I_2^{\alpha/\beta}\cdots I_n^{\alpha/\beta}$ can be analogously prepared by irradiating a different single transition.

More precisely, the operations of preparing a LPPS $I_{0z}I_1^{\alpha/\beta}I_2^{\alpha/\beta}\cdots I_n^{\alpha/\beta}$ before the final PFG can actually correspond to a series of unitary transformations instead of the specific pulse sequence in NMR. For example, the quantum circuit for preparing the LPPS $I_{0z}I_1^\alpha I_2^\alpha \cdots I_n^\alpha$ is shown in Fig. 1(a), consisting of $n+1$ single-qubit rotations and a $(n+1)$-qubit controlled-$R_{0y}$ gate that performs the operation $R_y$ on spin $I_0$ if all the states of $n$ controlling qubits are in $|0\rangle$. Here, $R_{iy} = e^{-iI_{iy}\pi/2}$ representing a $\pi/2$ rotation along the $y$ axis on spin $I_i$ and $R_{0y}^\dagger$ is its conjugated operator. The $(n+1)$-qubit controlled-$R_{0y}(\alpha)$ gate can be expressed as a propagator,

$$U_{control}(\alpha) = e^{-i\alpha I_{0y} I_1^\alpha I_2^\alpha \cdots I_n^\alpha}$$

$$= \begin{pmatrix} \cos(\alpha/2) & -\sin(\alpha/2) & & & \\ \sin(\alpha/2) & \cos(\alpha/2) & & & \\ & & 1 & & \\ & & & \ddots & \\ & & & & 1 \end{pmatrix}. \qquad (3)$$

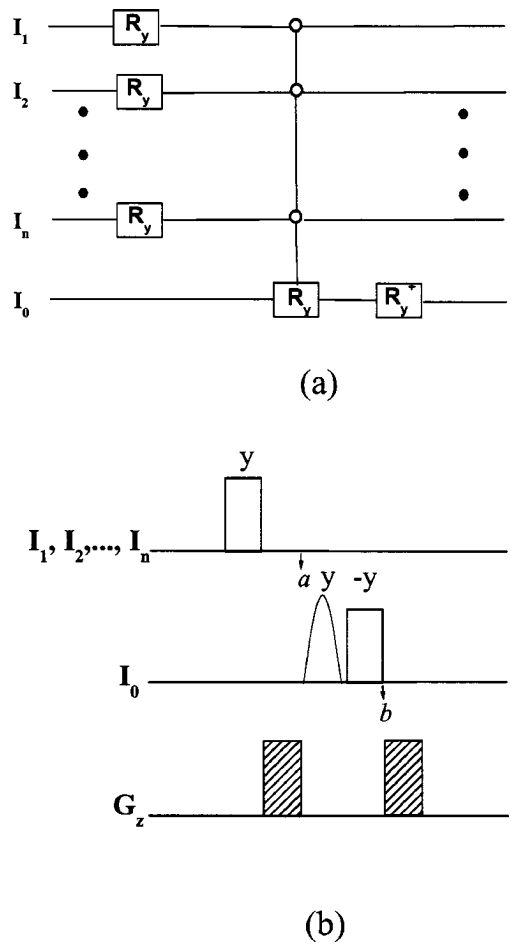

FIG. 1. Quantum circuit (a) and pulse sequence (b) for creating the LPPS $I_{0z}I_1^\alpha I_2^\alpha \cdots I_n^\alpha$ from thermal equilibrium. The controlled-$R_{0y}$ gate is performed conditional on all computational qubits $I_1, I_2, \ldots, I_n$ being in the $|0\rangle$ state represented by open circles. $R_y$ and $R_y^\dagger$ are described in the text. The first $\pi/2$ pulse is applied on all computational qubits $I_1, I_2, \ldots, I_n$ and the second $\pi/2$ pulse (denoted by the arc) represents the TSE on a single transition of the observer spin $I_0$. The PFGs ($G_z$) destroy all transverse magnetization.

Thus the quantum circuit in Fig. 1(a) yields an overall unitary propagator $U = R_{0y}^\dagger U_{control}(\pi/2) \prod_{i=1}^{n} R_{iy}$ that can transform thermal equilibrium $\rho_{eq}$ of Eq. (2) into the output state $\rho_f = U\rho_{eq}U^\dagger$. Since no zero-quantum transition is generated in the procedure, a consequent PFG eliminates all off-diagonal elements of $\rho_f$ and the remaining diagonal matrix is just equivalent to the LPPS $I_{0z}I_1^\alpha I_2^\alpha \cdots I_n^\alpha$. Thus we initialized the computational qubits into the pure state $I_1^\alpha I_2^\alpha \cdots I_n^\alpha$. Note that the state $I_{0z}I_1^\alpha I_2^\alpha \cdots I_n^\alpha$ is a truly mixed state, which is the tensor product of the observer spin operator $I_{0z}$ and the pure state $I_1^\alpha I_2^\alpha \cdots I_n^\alpha$. Each state of $I_{0z}I_1^{\alpha/\beta}I_2^{\alpha/\beta}\cdots I_n^{\alpha/\beta}$ can be simply obtained by modifying the state of the controlling qubits in the controlled-$R_{0y}$ gate according to the pure state $I_1^{\alpha/\beta}I_2^{\alpha/\beta}\cdots I_n^{\alpha/\beta}$. For example, in order to prepare the $I_{0z}I_1^\alpha I_2^\beta I_3^\beta$ state, the value of the control string in the quantum circuit should be $|0\rangle_1|1\rangle_2|1\rangle_3$. This can be in fact achieved by conjugating the controlled-$R_{0y}$ gate with NOT gates,[6] i.e., to sandwich the controlled-$R_{0y}$ gate between two NOT gates applied on those qubits being in the $I_i^\beta$ state in the desired LPPS.

Therefore, the quantum circuit in Fig. 1(a) can be effectively implemented by using a set of universal logic gates[4] which can be straightforwardly implemented in NMR.[29] In particular, conventional NMR spectroscopy techniques provide a more natural way to implement directly some multi-qubit logic gates,[29,30] such as a kind of crucial quantum controlled operations in quantum computing, just as described in Madi's multicomponent excitation scheme. When the molecule chosen in the experiment displays resolved $J$ couplings, due to no delays and refocusing schemes,[31] the direct implementation of a multi-qubit controlled-rotation gate by transition-selective pulses is simpler than that by standard $J$ coupling gates sandwiched between spin-selective pulses.[29,32] The low-power, long-duration TSE is, however, only an approximation of the logic gate and involves inevitably unwanted evolution under the internal Hamiltonian, so the entire Hamiltonian of the system should be considered to predict more precise details of the spin's evolution. Accordingly, in our experiment Madi's scheme for a LPPS with TSE is improved as an actual pulse sequence shown in Fig. 1(b), in which another PFG is inserted before the transition-selective pulse to avoid the unwanted evolution and the fast transverse relaxation effect during the long-duration TSE. As will be shown below, by analyzing theoretically the full dynamics of the excitation,[33] the pulse sequence in Fig. 1(b) is very efficient and reliable for preparing a LPPS on thermal equilibrium.

## B. Experimental demonstration and analyses

The physical system to demonstrate the above procedure was selected as the carbon-13 labeled alanine $NH_3^+ - C^\alpha H(C^\beta H_2) - C'O_2^-$ dissolved in $D_2O$ and operated on a Bruker ARX500 spectrometer with respect to transmitter frequencies of 500.13 MHz for $^1H$ and 125.77 MHz for $^{13}C$. The measured NMR parameters are listed in Table I. Due to its resolved scalar $J$ couplings to all other spins, $C^\alpha$ was chosen as the observer spin $I_0$. $C'$, $C^\beta$, and $H$ being directly joined with $C^\alpha$ are spins $I_1$, $I_2$, and $I_3$, respectively. The methyl protons were decoupled during the whole experiment. In the experiment, the transition-selective $\pi/2$ pulse on spin $I_0$ used to realize the controlled-$R_{0y}$ gate was *Gaussian* in shape and of 80 ms duration in order to achieve sufficient selectivity in the frequency domain without disturbing the nearest line.

TABLE I. Measured NMR parameters for alanine on a Bruker ARX500 spectrometer.

| Spin | $\nu$/Hz | $J_C$/Hz | $J_{C^\alpha}$/Hz | $J_{C^\beta}$/Hz | $J_H$/Hz |
|---|---|---|---|---|---|
| $C'$ (1) | $-4320$ | | 34.94 | $-1.2$ | 5.5 |
| $C^\alpha$ (0) | 0 | 34.94 | | 53.81 | 143.21 |
| $C^\beta$ (2) | 15 793 | $-1.2$ | 53.81 | | 5.1 |
| $H$ (3) | 1550 | 5.5 | 143.21 | 5.1 | |

We now turn to analyze theoretically the TSE. In the weak coupling limit $2\pi J_{ij} \ll |\omega_i - \omega_j|$, the internal Hamiltonian of an arbitrary $(n+1)$-spin system in a large static magnetic field can be expressed as[34]

$$H_{int} = -\sum_{i=0}^n \omega_i I_{iz} + 2\pi \sum_{i<j}^n \sum_{i=0}^n J_{ij} I_{iz} I_{jz}, \quad (4)$$

where $J_{ij}$ is the coupling constant between spins $i$ and $j$. The external Hamiltonian for the applied RF field with the amplitude $B_1$ has the form

$$H_{ext} = \sum_{i=0}^n e^{-i(\omega_{rf}t+\phi)} \Omega_i I_{ix} e^{i(\omega_{rf}t+\phi)}, \quad (5)$$

where $\omega_{rf}$ is the transmitter's angular frequency, $\phi$ the initial phase and $\Omega_i = \gamma_i B_1$ the Rabi frequencies controlled experimentally by adjusting the RF powers. For simplicity, assuming that the amplitude $B_1$ of the applied RF field is constant over the duration of the pulse, without considering the relaxation effects, the total Hamiltonian $H = H_{int} + H_{ext}$, whose time dependence can be removed by transforming into a new frame rotating at the frequency $\omega_{rf}$. The time-independent effective Hamiltonian in the new rotating frame[34] is then given by

$$H_{eff} = \sum_{i=0}^n (\omega_{rf} - \omega_i) I_{iz} + 2\pi \sum_{i<j}^n \sum_{i=0}^n J_{ij} I_{iz} I_{jz} + \sum_{i=0}^n \Omega_i I_{iy}. \quad (6)$$

In Eq. (6), $\phi$ is taken to be $\pi/2$. If $\omega_{rf} = \omega_0 - \pi\Sigma_{k=1}^n J_{0k}$ matches the $|000..0\rangle \leftrightarrow |100...0\rangle$ transition of the multiplet components of spin $I_0$, it is shown in the Appendix that if $\Omega_0 \ll 2\pi \min(J_{0i})$, the full evolution under the effective Hamiltonian $H_{eff}$ can be approximately decomposed as

$$U_0(\alpha) = e^{-i\alpha H_{eff}/\Omega_0} \approx U_{control}(\alpha) U_z(\alpha) \quad (7)$$

by generalizing the method in Ref. 33 into general cases with arbitrary controlled rotation on multispin systems, where $U_z(\alpha)$ denotes an additional conditional phase factor. Hence, the propagator of the effective Hamiltonian of the TSE over the duration of the pulse $\tau = \alpha/\Omega_0$ is equivalent to $U_{control}(\alpha)$ of Eq. (3) up to conditional phases as long as $\Omega_0 \ll 2\pi \min(J_{0i})$. Figure 2 shows the numerical simulations of the gate fidelity[35] $F(U_0, U_2) = |Tr(U_0^\dagger U_2)/N|^2$ under different exciting angles $\alpha$ for (a) a 3-spin homonuclear system and (b) a 4-spin heteronuclear system on the carbon-13 labeled alanine. Here, $U_2(\alpha)=U_{\text{control}}(\alpha)U_z(\alpha)$ is the approximate propagator whereas $U_0(\alpha)$ is the exact one under the effective Hamiltonian $H_{\text{eff}}$. We take a quality factor $Q=1-F$. The figure manifests that when $\Omega_0/\pi J_{01}$ is small enough (e.g., <0.4), the gate fidelities can reach above the order of 0.99, e.g., for the transition-selective pulse employed in our experiment, $\Omega_0/\pi J_{01}\approx 0.179$ for $\tau=80$ ms and theoretically the gate fidelity $F(U_0,U_2)\approx 0.999$. It can be seen from Fig. 2 that the gate fidelities in a 4-spin system are higher than that in a 3-spin system as a result of the fact $J_{03} > J_{01}$, $J_{02}$ for the alanine molecule.

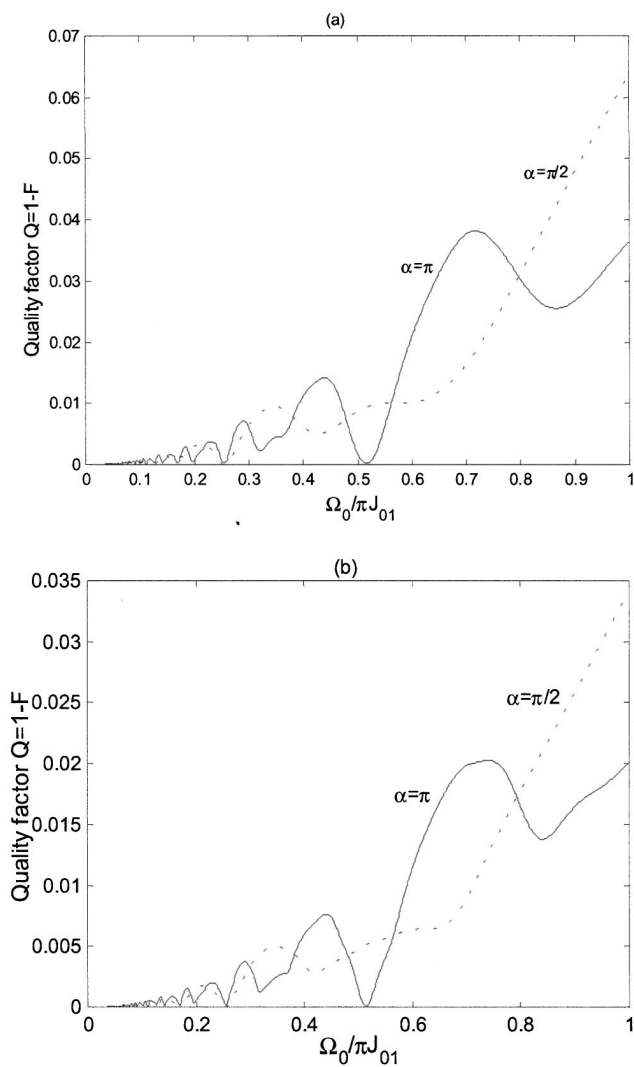

FIG. 2. The numerical simulations of the quality factor of gate $Q=1-F(U_0,U_2)$ in (a) a 3-spin homonuclear system and (b) a 4-spin heteronuclear system with the alanine molecule in different angles $\alpha=\pi/2$ (denoted by the dotted line) and $\pi$ (denoted by the solid line). $U_0$ and $U_2$ are described in the text.

Due to the existence of the additional conditional phase factor $U_z(\alpha)$, the single TSE is incorrectly equated to a simple controlled gate $U_{\text{control}}(\alpha)$ in the general cases. For example, if the transition-selective pulse is directly applied after $n\,R_{iy}$ in our experiment, the spin evolution under the internal Hamiltonian during the pulse length of 80 ms strongly affects the intended experiment which produces a big phase factor. However, the goal of $n\,R_{iy}$ operations is to transform the longitudinal magnetizations of $n$ computational spins of thermal equilibrium into the transverse ones which can be eliminated by the PFG at the end of the sequence. Consequently, another PFG is inserted after $n\,R_{iy}$ operations to reach the goal first. The sequence is in fact reduced to a selective pulse on the initial state $I_{0z}$ since $\rho_f = U_2(\alpha)I_{0z}U_2^{\dagger}(\alpha) = U_{\text{control}}(\alpha)I_{0z}U_{\text{control}}^{\dagger}(\alpha)$ with the commutivity between the phase factor $U_z(\alpha)$ and $I_{0z}$. Hereby, the effects of the conditional phases are canceled out and the transition-selective pulse in practice implements the controlled-gate $U_{\text{control}}(\alpha)$ on those input states which commute with $U_z(\alpha)$. Alternatively, one can first apply the transition-selective pulse on thermal equilibrium and then $n+1\,R_{iy}$ operations on all spins followed by a PFG to produce a LPPS.

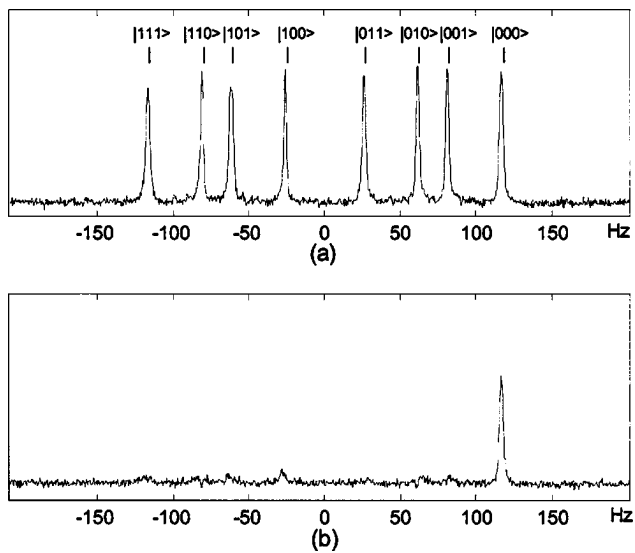

FIG. 3. Experimental spectra (in arbitrary units) for (a) the reference spectrum in thermal equilibrium and (b) the LPPS $I_{0z}I_1^{\alpha}I_2^{\alpha}\cdots I_n^{\alpha}$. All spectra were obtained by a spin-selective $(\pi/2)_y^0$ pulse on the observer spin $I_0$.

The LPPS $I_{0z}I_1^{\alpha}I_2^{\alpha}I_3^{\alpha}$ obtained experimentally is shown in Fig. 3, along with the reference spectrum of $C^{\alpha}$. Figure 3(b) shows that only one transition of spin $I_0$ corresponding to $I_1^{\alpha}I_2^{\alpha}I_3^{\alpha}(|000\rangle\langle 000|)$ is retained, indicating that the computational qubits were successfully prepared into the logic state $|000\rangle$. Meanwhile, by a comparison of two spectral signals in Fig. 3, the signal intensity of the labeled PPS was reduced to about 80%. To roughly estimate the intensity decay of the observer spin during the single TSE,[36] we obtain $M(t)/M(0)=e^{-\tau/T_2}=0.82$ where $T_2$ is the spin–spin relaxation time of spin $I_0$ measured to be 0.41 s, $M(0)$ and $M(t)$ are the intensity of the spin before and after the excitation, respectively. Therefore, the main cause for the signal decay during the preparation of the LPPS is the relaxation effect of the low-power, long-duration TSE. Besides, the sources of experimental errors are also due to inhomogeneity of RF fields and static magnetic fields, the imperfections of the spin- and transition-selective excitations, etc.

## III. NMR SPECTRAL IMPLEMENTATION OF THE BERNSTEIN–VAZIRANI ALGORITHM

The Bernstein–Vazirani (BV) problem[11] can be expressed as follows: Suppose that a black box *oracle* computes a function $f:\{0,1\}^n \to \{0,1\}$ where $f_a(x) = a \cdot x = (a_1 \wedge x_1) \oplus (a_2 \wedge x_2) \oplus \ldots \oplus (a_n \wedge x_n)$, and $a, x \in \{0,1\}^n$, $a_i$ and $x_i$ are the $i$th qubits of $a$ and $x$, and consider the goal of determining $a$. Classically, it would require $n$ queries of the *oracle* to determine $a$ with certainty. In 1993, Bernstein and Vazirani[11] first gave a quantum algorithm to solve this problem with two queries, which was slightly improved to comprise a single query.[37,38] The scheme for BV algorithm[37] is illustrated in Fig. 4. The quantum *oracle* is enacted by a unitary operation $U_f$ on the computational basis

$$U_f : |x\rangle|y\rangle \to |x\rangle|y \oplus f(x)\rangle, \tag{8}$$

where $|x\rangle$ is the database register, $\oplus$ denotes addition modulo 2, and a single qubit $|y\rangle$ is the oracle qubit. The whole procedure can be described as

$$|0\rangle^n \otimes \frac{|0\rangle - |1\rangle}{\sqrt{2}} \xrightarrow{H^{(n)}} \frac{1}{\sqrt{2^n}} \sum_{x=0}^{2^n-1} |x\rangle \otimes \frac{|0\rangle - |1\rangle}{\sqrt{2}} \xrightarrow{U_f} \frac{1}{\sqrt{2^n}} \sum_{x=0}^{2^n-1} (-1)^{a \cdot x} |x\rangle \otimes \frac{|0\rangle - |1\rangle}{\sqrt{2}} \xrightarrow{H^{(n)}} \frac{1}{\sqrt{2^n}} \sum_{x,y=0}^{2^n-1}$$

$$\times (-1)^{x \cdot (a \oplus y)} |y\rangle \otimes \frac{|0\rangle - |1\rangle}{\sqrt{2}} \equiv |a\rangle \otimes \frac{|0\rangle - |1\rangle}{\sqrt{2}}. \tag{9}$$

Measuring the $n$-qubit database register yields the value of $a$ with probability one. Although entanglement can appear in the black box by the oracle operation $U_f$ in the course of computation, no entanglement is involved at any step as the algorithm starts with a product state.[35,37] Therefore, our experiment focuses on the refined version[39] in which the oracle is redesigned as an $n$-qubit unitary transformation $U_a$,

$$U_a : |x\rangle \to (-1)^{a \cdot x} |x\rangle, \tag{10}$$

instead of the $(n+1)$-qubit unitary transformation $U_f$. The gate $U_a$ can be rewritten as a direct product of single qubit operators $U_a = \Pi_j (\sigma_{jz})^{a_j}$, using the definition $(\sigma_{j\eta})^0 = E_j$. The action of the whole algorithm $U_{BV} = H^{(n)} U_a H^{(n)}$ $= \Pi_j (\sigma_{jx})^{a_i}$ which can be implemented by applying $\pi$ pulses on spins $I_i$ if $a_i = 1$. The simple operation trivially gives the result of the BV problem.

For the LPPS $I_{0z} I_1^\alpha I_2^\alpha \cdots I_n^\alpha$, the algorithm is applied to the computational qubits $I_1, I_2, \ldots, I_n$, while the observer spin $I_0$ is not directly involved into the quantum computation. After a spin-selective readout pulse $(\pi/2)_y^0$, one can detect the magnetization of spin $I_0$,[34]

$$M_0 = \text{Tr}[I_0^+ (I_{0x} \otimes U_{BV}(I_1^\alpha I_2^\alpha \cdots I_n^\alpha) U_{BV}^\dagger)]$$
$$= \text{Tr}[I_0^+ (I_{0x} \otimes \rho_{out})] = \rho_{11} + \rho_{22} + \cdots + \rho_{nn}, \tag{11}$$

where $I_0^+ = I_{0x} + iI_{0y}$ and $\rho_{ii}$ related to the database $|i-1\rangle (i = 1, 2, \ldots, N = 2^n)$ indicates the $i$th diagonal element of the output density matrix $\rho_{out} = U_{BV}(I_1^\alpha I_2^\alpha \cdots I_n^\alpha) U_{BV}^\dagger$, corresponding to each of $2^n$ individual spectral resonance lines in the spectrum of spin $I_0$, respectively. If all of $J_{0i} > 0$, $\rho_{11}$ is the transition at the lowest frequency and $\rho_{NN}$, the one at the highest frequency as illustrated in Fig. 3(a). That is, the results of projective measurements under the computational basis of qubits can be obtained from the split peak positions in the spectra. Theoretically, $\rho_{out} = |a\rangle\langle a|$ for the BV algorithm denotes that the state of the $n$ computational qubits is in the state $|a\rangle$, i.e., $\rho_{ii} = 1$ for $|i-1\rangle = |a\rangle$ while $\rho_{ii} = 0$ for all other $|i-1\rangle \neq |a\rangle$ in Eq. (11). So only the transition line of spin $I_0$ relevant to the state $|a\rangle$ is retained in the spectrum of spin $I_0$. The "answer" of the computation is intuitively identified by the spectrum of spin $I_0$. From Eq. (11), we can see that if the "answers" of some computational problems are only relevant to the information of the diagonal elements of the output density matrix, or the results of projective measurements under the computational basis of qubits are only needed in the algorithms, the readout of spectroscopical mapping is favorable.

The experimental results of the BV algorithm on the LPPS prepared above are presented in Fig. 5, which are consistent with the theoretical predictions. Detections of all states were achieved by a spin-selective $(\pi/2)_y^0$ pulse, followed by acquisition and Fourier transformation of the signal.[34] The spectrum of spin $I_0$ carries the results of the quantum computations. Due to the algorithm implemented by only several spin-selective $\pi$ pulses on computational qubits, the experimental errors are on the same order with the LPPS.

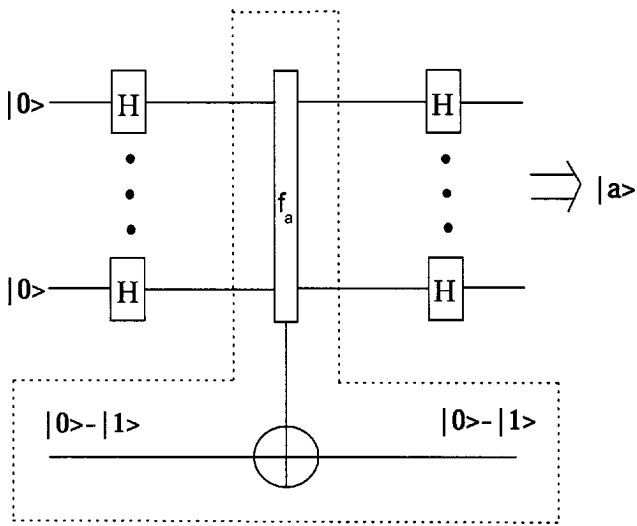

FIG. 4. Quantum circuit for BV algorithm. $H$ represents the Hadamard gate. The unitary operation $U_f$ in the dashed box represents a black box query, which can be simplified as another unitary transformation $U_a$ in the refined version (see the text).

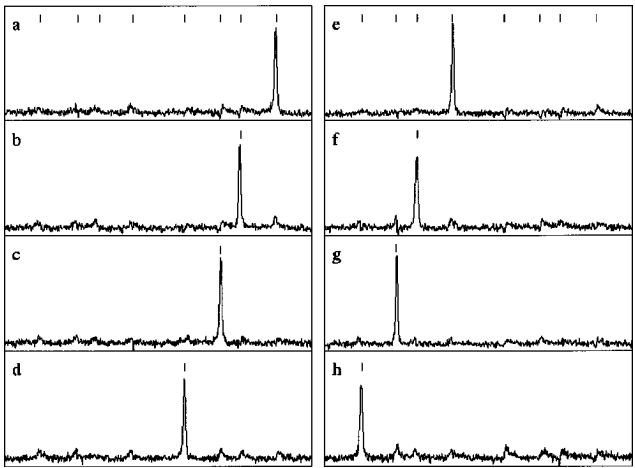

FIG. 5. Experimental spectra of implementing the quantum algorithm to solve different BV problems for (a) $a=000$, (b) $a=001$, (c) $a=010$, (d) $a=011$, (e) $a=100$, (f) $a=101$, (g) $a=110$ and (h) $a=111$.

## IV. CONCLUSION

In summary, by introducing a quantum circuit into the preparation of the LPPS, we have experimentally implemented a method of preparing a LPPS by the multiplet-component excitation on a 4-qubit NMR quantum processor. The controlled-rotation gate is of central importance in quantum computing. We choose the direct implementation of the TSE scheme, and by analyzing theoretically its entire Hamiltonian, a transition-selective pulse can yield a controlled-rotation gate up to phase factors with high fidelity when $\Omega_0 \ll 2\pi \min(J_{0i})$. So it is especially efficient for the present preparation scheme of the LPPS on thermal equilibrium. But the long-duration TSE greatly causes the signal decay due to relaxation effects and requires a very suitable sample molecule in which one spin has well-resolved $J$ couplings to others. Moreover, based on "spectral implementation," a 3-qubit BV algorithm has been experimentally demonstrated on such a LPPS. In the spectrum of spin $I_0$ detected at the end of experiments, the label of the highest peak gives the same answer as the projective measurement under the computational basis of qubits, which is good agreement with theory. The introduction of the observer spin $I_0$ provides a convenient way to readout the results of the computation efficiently and reliably for certain specific computational problems.

## ACKNOWLEDGMENTS

The authors thank Xiaodong Yang, Hanzeng Yuan, and Xu Zhang for help in the course of experiments. This work is supported by the National Fundamental Research Program (2001CB309300).

## APPENDIX: DECOMPOSITION OF THE ENTIRE PROPAGATOR FOR A SINGLE-TRANSITION EXCITATION

For a TSE, a basic requirement in experiment is $(\gamma_0/2\pi)B_1| < |\min(J_{0i})|$, thus the off-resonance effects due to the $I_{iy}$ terms in $H_{\text{eff}}$ of Eq. (6) can be ignored in a weak-coupled system.[33] Consequently, we deal with $H_{\text{eff}}^0 = H_{\text{eff}} - \Sigma_{k=1}^n \Omega_k I_{ky}$ instead of $H_{\text{eff}}$ with a very good approximation. Consider a 3-spin system and let

$$a_i = \frac{\omega_0 - \omega_i}{\Omega_0}, \quad b_i = \frac{\pi J_{0i}}{\Omega_0}, \quad \text{and} \quad c_1 = \frac{\pi J_{12}}{\Omega_0} \quad (i=1,2). \tag{A1}$$

One gets from Eq. (6),

$$H_{\text{eff}}^0 / \Omega_0 |_{n=2} = P + Q_{--} I_1^\beta I_2^\beta + Q_{+-} I_1^\alpha I_2^\beta + Q_{-+} I_1^\beta I_2^\alpha + Q_{++} I_1^\alpha I_2^\alpha, \tag{A2}$$

where

$$P = \sum_{i=1}^{2} (a_i - b_1 - b_2) I_{iz} + 2c_1 I_{1z} I_{2z},$$

$$Q_{--} = -2(b_1+b_2) I_{0z} + I_{0y},$$

$$Q_{-+} = -2b_1 I_{0z} + I_{0y},$$

$$Q_{+-} = -2b_2 I_{0z} + I_{0y},$$

$$Q_{++} = I_{0y}. \tag{A3}$$

Note that the five terms in Eq. (A2) commute each other. $P$ is a diagonal matrix and $H_{trn} = Q_{++} I_1^\alpha I_2^\alpha$. However, $Q_{--} I_1^\beta I_2^\beta$, $Q_{-+} I_1^\beta I_2^\alpha$, and $Q_{+-} I_1^\alpha I_2^\beta$ are not diagonal and can be readily diagonalized[33] since

$$Q_{--} = -e^{-i\theta_1 I_{0x}} I_{0z} e^{i\theta_1 I_{0x}} \sqrt{1+4(b_1+b_2)^2},$$

$$\theta_1 = \arctan(1/(2b_1+2b_2)), \tag{A4}$$

and the similar expressions for $Q_{-+}$ with $\theta_2 = \arctan(1/(2b_1))$ and $Q_{+-}$ with $\theta_3 = \arctan(1/(2b_2))$. Therefore, we can evaluate the difference between two unitary propagators without and with this transformation $U_T = e^{i\theta_3 I_{0x} I_1^\alpha I_2^\beta} e^{i\theta_2 I_{0x} I_1^\beta I_2^\alpha} e^{i\theta_1 I_{0x} I_1^\beta I_2^\beta}$, i.e., $U_1 = e^{-i\alpha H_{\text{eff}}^0/\Omega_0}$ and $U_2 = U_T U_1 U_T^\dagger$ by calculating the gate fidelity between $U_1$ and $U_2$, which can be defined by

$$\begin{aligned} F(U_1,U_2) &= |\text{Tr}(U_1^\dagger U_2)/N|^2 = |\text{Tr}(e^{i\alpha Q_- I_1^\beta I_2^\beta} e^{i\alpha Q_{+-} I_1^\alpha I_2^\beta} e^{i\alpha Q_{-+} I_1^\beta I_2^\alpha} \\ &\quad \cdot e^{i\alpha\sqrt{1+4(b_1+b_2)^2} I_{0x} I_1^\beta I_2^\beta} e^{i\alpha\sqrt{1+4b_2^2} I_{0z} I_1^\alpha I_2^\beta} e^{i\alpha\sqrt{1+4b_1^2} I_{0z} I_1^\beta I_2^\alpha})/8|^2 \\ &= \Bigg| 1 - \frac{1}{2^2}\bigg[\sin^2\!\left(\frac{\alpha}{2}\sqrt{1+4(b_1+b_2)^2}\right)\!\left(1 - \frac{2(b_1+b_2)}{\sqrt{1+4(b_1+b_2)^2}}\right) + \sin^2\!\left(\frac{\alpha}{2}\sqrt{1+4b_2^2}\right)\!\left(1 - \frac{2b_2}{\sqrt{1+4b_2^2}}\right) \\ &\quad + \sin^2\!\left(\frac{\alpha}{2}\sqrt{1+4b_1^2}\right)\!\left(1 - \frac{2b_1}{\sqrt{1+4b_1^2}}\right)\bigg]\Bigg|^2, \end{aligned} \quad (A5)$$

where $N$ is the dimension of the Hilbert space. In the square brackets, the second factor of each term is much less than 1 if $\Omega_0 \ll 2\pi\min(J_{0i})$, e.g.,

$$1 - \frac{2b_1}{\sqrt{1+4b_1^2}} \leq \frac{1}{4b_1^2} = \frac{\Omega_0^2}{(2\pi J_{01})^2} \ll 1$$

so that $F(U_1,U_2)\to 1$. Then it follows

$$e^{-i\alpha H_{\text{eff}}} \approx e^{-i\alpha H_{\text{eff}}^0} \approx U_{\text{control}}(\alpha) U_z(\alpha) \quad (A6)$$

with

$$U_z(\alpha) = e^{-i\alpha P} e^{i\alpha\sqrt{1+4(b_1+b_2)^2} I_{0z} I_1^\beta I_2^\beta} \\ \times e^{i\alpha\sqrt{1+4b_2^2} I_{0z} I_1^\alpha I_2^\beta} e^{i\alpha\sqrt{1+4b_1^2} I_{0z} I_1^\beta I_2^\alpha}. \quad (A7)$$

Here, $U_{\text{control}}(\alpha) = e^{-i\alpha H_{trn}}$ is equal to Eq. (3). Hence, the propagator of the effective Hamiltonian of the TSE is equivalent to $U_{\text{control}}(\alpha)$ of Eq. (3) up to a conditional phase factor $U_z(\alpha)$. For any $n$-spin system, the propagator has approximately the analogous decomposition, i.e., except for the transition component irradiated, additional phase factors are yielded for other components of this spin and other spins if $\Omega_0 \ll 2\pi\min(J_{0i})$. In a two-spin system, the exact controlled-rotation gate can be implemented at certain specific power levels, i.e., $F(U_1,U_2)=1$ for certain specific power levels, while for $n>2$, $F(U_1,U_2)\to 1$ at a high enough precision.